\documentclass[twocolumn,showpacs,prl]{revtex4}
\usepackage{graphics}
\usepackage{epsfig}
\usepackage{dcolumn}
\usepackage{amsmath}

\begin{document}

\title{Anomalous $c$-axis transport in layered metals}
\author{D. B. Gutman and D. L. Maslov}
\date{\today}

\begin{abstract}
Transport in metals with strongly anisotropic single-particle
spectrum is studied. Coherent band transport in all directions, described by the standard Boltzmann equation, is
shown to withstand both elastic and inelastic scattering as long as $E_F\tau\gg 1$. A model of
phonon-assisted tunneling via resonant states located in between the layers is suggested to
explain a non-monotonic temperature dependence of the $c$-axis resistivity observed in experiments.
\end{abstract}

\affiliation{Department of Physics, University of Florida, Gainesville,
FL 32611, USA}
 \maketitle

Electron transport in layered materials exhibits a number of unusual
properties. The most striking example is a qualitatively different behavior
of the in-plane $\left( \rho _{ab}\right) $ and out-of-plane ($\rho _{c})$
resistivities: whereas the temperature dependence of $\rho _{ab}$ is
metallic-like, that of $\rho _{c}$ is either insulating-like or even
non-monotonic. At the level of non-interacting electrons, layered systems
are metals with strongly anisotropic Fermi surfaces. A commonly used model
is free motion along the planes and nearest-neighbor hopping between the
planes:
\begin{equation}
\varepsilon _{\mathbf{k}}=\mathbf{k}_{||}^{2}/2m_{ab}+2J\left( 1-\cos
k_{\perp }d\right) ,  \label{spectrum}
\end{equation}
where $\mathbf{k}_{||}$ and $k_{\perp }$ are in the in-plane and $c$-axis
components of momentum, respectively, $m_{ab}$ is the in-plane mass, and $d$
is lattice constant in the $c$-axis direction. For the strongly anisotropic
case ($J\ll E_{F}),$ the equipotential surfaces are ``corrugated cylinders''
(see Fig.1).

If the Hamiltonian consists of the band motion with spectrum (\ref{spectrum}%
) and the interaction of electrons with potential disorder as well as with
inelastic degrees of freedom, e.g., phonons, the Boltzmann equation predicts
that the conductivities are given by
\begin{equation}
\sigma _{ab}^{B}=e^{2}\nu \langle v_{a}v_{b}\tau _{\text{tr}}\rangle
,\;\sigma _{c}^{B}=4e^{2}\nu J^{2}d^{2}\langle \sin ^{2}\left( k_{\perp
}d\right) \tau _{\text{tr}}\rangle ,  \label{sigma_b}
\end{equation}
where $\langle \dots \rangle $ denotes averaging over the Fermi surface and
over the thermal (Fermi) distribution, $\nu =m_{ab}/\pi d$ is the density of
states, and $\tau _{\text{tr}}$ is the transport time, resulting from all
scattering processes (we set $\hbar =k_{B}=1)$. If $\tau _{\text{tr}}$
decreases with the temperature, \emph{both }$\sigma _{ab}$ and $\sigma _{c}$
are expected to decrease with $T$ as well. This is not what the experiment
shows.


\begin{figure}[tbp]
\includegraphics[angle=0,width =0.3\textwidth]{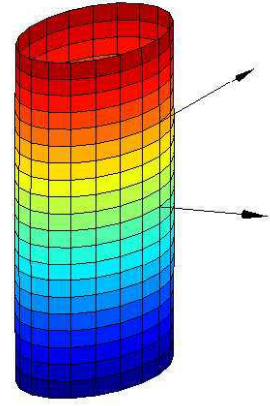}
\caption{Fermi surface corresponding to Eq.(\ref{spectrum})
 with Fermi velocity vectors at two different points.}
\end{figure}

The $c$-axis puzzle received a lot of attention in connection to the HTC
materials~\cite{ginsberg}, and a non-Fermi-liquid nature of these materials was suggested to be responsible for the anomalous $%
c$-axis transport \cite{anderson}. However, other materials, such as
graphite \cite{graphite}, TaS$_{2}$ \cite{frindt}, Sr$_{2}$RuO$_{4}$ \cite
{srruo}, organic metals \cite{organics}, etc., behave as canonical Fermi liquids in all aspects but the $c$%
-axis transport. This suggests that the origin of the effect is
not related to the specific properties of HTC compounds but
common for all layered materials. A large number of models were
proposed to explain the $c$-axis puzzle. Despite this variety, most authors
seem to agree on that the coherent band transport in the
$c$-axis direction is destroyed. Although there is no agreement as
to what replaces the band transport in the ''incoherent'' regime,
the most frequently discussed mechanisms include incoherent
tunneling between the layers, assisted by either out-of-plane
impurities \cite{sauls,levin,peter,abrikosov_res} or by coupling
to dissipative environment \cite{leggett}, and polarons
\cite {polaron_schofield,polaron_mckenzie}. 

The message of this
Letter is two-fold. First, we observe that neither elastic or
inelastic (electron-phonon) scattering can destroy band transport
even in a strongly anisotropic metal as long as the familiar
parameter $E_{F}\tau $ is large. Nothing happens to the Boltzmann
conductivities in Eq.(\ref{sigma_b}) except for $\sigma _{c}^{B}$
becoming very small at high temperatures so that other mechanisms,
not included in Eq.(\ref{sigma_b}), dominate transport. This observation is in agreement with
recent experiment \cite{singleton} where a coherent feature (angle-dependent
magnetoresistance) was observed in a supposedly incoherent regime. Second, we
propose phonon-assisted tunneling through resonant impurities as the
mechanism competing with the band transport. As such tunneling
provides an additional channel for transport, the total
conductivity is \cite{levin}
\begin{equation}
\sigma _{c}=\sigma _{c}^{B}+\sigma _{\text{res}},  \label{cond}
\end{equation}
where $\sigma _{\text{res}}$ is the resonant-impurity contribution. Because $%
\sigma _{\text{res}}$ increases with the temperature, the band
channel is short-circuited by the resonant one at high enough
temperatures\cite{Palevskii}. Accordingly, $\sigma _{c}$ goes through a minimum at
a certain temperature (and $\rho _{c}=\sigma _{c}^{-1}$ goes
through a maximum). 
We consider
phonon-assisted tunneling through a wide band of resonant levels
distributed uniformly in space. We show that the
non-perturbative (in the electron-phonon coupling) version of this
theory is in a quantitative agreement with the experiment on
Sr$_{2}$RuO$_{4}$ \cite{srruo}. Due to a similarity  between 
phonon-assisted tunneling  and  other  problems, in which interaction leads 
to the formation  of  a cloud surrounding the electron
(such as polaronic effect and zero bias anomaly),
 many ideas put forward  earlier  
\cite{sauls,levin,peter,abrikosov_res,leggett,polaron_schofield,polaron_mckenzie}
agree  with our picture.
Nevertheless, we believe that only a combination of resonant impurities 
and electron-phonon interaction 
solves the puzzle of  $c$-axis resistivity  and provides a microscopic
theory for some of the mechanisms considered in prior work.  We
begin with the discussion of the breakdown (or lack of it thereof)
of the Boltzmann equation.

One may wonder whether the band transport along the $c$-axis
breaks down because the Anderson localization transition occurs
in the $c$-direction whereas the in-plane
transport remains metallic. This does not happen, however, because
an electron, encountering an obstacle for motion along the
$c$-axis, moves quickly to another point in the plane, where such
an obstacle is absent. More formally, it has been shown the
Anderson transition occurs only simultaneously in all directions
\cite{woelfle_loc,lee,dupuis} and only if $J$ is {\em exponentially} smaller than $1/\tau$.
 Therefore, localization cannot explain
the observed behavior.

Refs.\cite{kumar,mckenzie} suggested an idea of the ``coherent-incoherent
crossover''. It implies that the coherent band motion breaks down if
electrons are scattered faster than they tunnel between adjacent layers,
i.e., if $J\tau \ll 1.$ Consequently, the current in the $c$-direction is
carried via incoherent hops between conducting layers.
It was noted by a number of authors that the assumption about incoherent
nature of the transport does not, by itself, explain the difference in
temperature dependences of $\sigma _{ab}$ and $\sigma _{c}$ \cite
{mckenzie,ioffe}: due to conservation of the in-plane momentum, $\sigma _{c}$
is proportional to $\tau $ both in the coherent and incoherent regimes.
%
Nevertheless, an issue of the ``coherent-incoherent crossover'' poses a
fundamentally important question: can scattering destroy band transport only in
some directions, if the spectrum is
anisotropic enough \cite{chaikin}? We argue here that this is not the case.

Since we have already ruled out elastic scattering, this leaves inelastic
one as a potential culprit. 
We focus on the case of the electron-phonon interaction as a source of
inelastic scattering. For an isotropic metal, the quantum kinetic equation
is derived from the Keldysh equations of motion for the Green's function via
the Prange-Kadanoff procedure \cite{rammer} for any strength of the
electron-phonon interaction.
In this Letter, we apply the Prange-Kadanoff theory to metals with
strongly anisotropic Fermi surfaces, such as the one in Fig. 1. We show
that, exactly as in the isotropic case, the Boltzmann equation holds its
standard form as long as $E_{F}\tau _{\text{e-ph}}\gg 1$. Since this form
does not change between coherent ($J\tau _{\text{\textrm{e-ph}}}\gg 1$)
and incoherent ($J\tau _{\text{e-ph}}\ll 1$) regimes, it means that the
coherent-incoherent crossover is, in fact, \emph{absent}.

We adopt the standard Fr{\"{o}}lich Hamiltonian for the
deformation-potential interaction with longitudinal acoustic phonons ($%
\omega _{q}=sq)$
\begin{eqnarray}&&
\hspace{-0.8cm}
H\!=\!\!\sum_{\mathbf{k}}\!\epsilon _{\mathbf{k}}a_{\mathbf{k}%
}^{\dagger }a_{\mathbf{k}}^{{}}\!+\!\sum_{\mathbf{q}}\!\omega _{q}b_{\mathbf{q}%
}^{\dagger}b_{\mathbf{q}}^{{}}  \label{frolich} 
\!+\!
\sum_{\mathbf{k,q}}\!g_q\!\sqrt{%
\omega _{q}}a_{\mathbf{k+q}}^{\dagger}a_{\mathbf{k}}^{{}}\!\left(\!b_{\mathbf{q
}}^{{}}\!+\!b_{-\mathbf{q}}^{\dagger}\!\right).
\end{eqnarray}
Since tunneling matrix elements are much more sensitive to the increase in
the inter-plane distance than the elastic moduli, the anisotropy of phonon
spectra in layered materials, albeit significant, is still weaker than the
anisotropy of electron spectra (see, e.g., Ref. \cite{elastic}). Therefore,
we treat phonons in the isotropic approximation, and assume that the \emph{%
magnitude} of the Fermi velocity is larger than the speed of sound $s.$

For a
static and uniform electric field, the Keldysh component of the electron's
Green function satisfies the Dyson equation 
\begin{eqnarray}
&&\hat{L}G^{K}+\frac{i}{2}\left( [\mathrm{Re}\Sigma ^{R},\otimes
G^{K}]_{-}+[\Sigma ^{K},\otimes \mathrm{Re}G^{K}]_{-}\right)  \notag \\
&=&\frac{1}{2}\left( [\Sigma ^{K},\otimes A]_{+}-[\Gamma ,\otimes
G^{K}]_{+}\right) \,.  \label{e10}
\end{eqnarray}
Here 
$\hat{L}=\left( \partial _{t}+\mathbf{v\cdot \nabla }_{\mathbf{R}}+e\mathbf{%
E\cdot \nabla }_{\mathbf{k}}\right) $ 
is the Liouville operator, $A=i(G^{R}-G^{A})$ is the spectral function, $%
\Gamma =i\left( \Sigma ^{R}-\Sigma ^{A}\right) $, and $\otimes $ denotes the
convolution in space and time. 
 Thanks to the Migdal theorem, the self-energy does not depend
on electron's dispersion $\xi _{{\bf k}}\equiv \varepsilon _{{\bf k}}-E_{F},$ and Eq.(%
\ref{e10}) can be integrated over $\xi _{{\bf k}}.$ This results in an equation
\begin{equation}
\hat{L}g^{K}+\frac{i}{2}[\mathrm{Re}\Sigma ^{R},g^{K}]_{-}=2i\Sigma ^{K}-%
\frac{1}{2}[\Gamma ,g^{K}]_{+}\,\,\,  \label{z1}
\end{equation}
for the ``distribution function''
\begin{equation}
g^{K}(\epsilon ,\hat{n})=\frac{i}{\pi }\int G^{K}(\epsilon ,\xi _{{\bf k}},\hat{n}%
)d\xi _{{\bf k}}\,,
\end{equation}
where $\hat{n}=\mathbf{v}_{\mathbf{k}}/\left| \mathbf{v}_{\mathbf{k}}\right|
$ is a local normal to the Fermi surface.

We consider a linear \emph{dc } response, when the self-energy is needed
only at equilibrium. Within the Migdal theory, the Matsubara self-energy is
given by a single diagram
\begin{equation*}
\Sigma (\epsilon ,\hat{n})=-\int \frac{d\omega }{2\pi }\int \frac{d^{3}q}{%
\left( 2\pi \right) ^{3}}g^{2}\left( q\right) G(\epsilon -\omega ,\mathbf{k}-%
\mathbf{q})D(\omega ,q)\,,
\end{equation*}
where the dressed phonon propagator
\begin{equation*}
D^{-1}=D_{0}^{-1}-g^{2}\Pi
\end{equation*}
is expressed through bare one
\begin{equation*}
D_{0}(\omega ,q)=-s^{2}q^{2}/\left( \omega ^{2}+s^{2}q^{2}\right)
\end{equation*}
and polarization operator $\Pi $ which, for $E_{F}>2J,$ is given by its 2D
form
\begin{equation*}
\Pi (\omega ,q)=-\nu \left( 1-|\omega |/\sqrt{v_{F}^{2}q_{\parallel
}^{2}+\omega ^{2}}\right) .
\end{equation*}
We assume that the electron-phonon vertex decays on some scale $k_{D}$
shorter than Fermi momentum ($k_{D}\ll k_{F}$). This assumption allows one
to linearize the dispersion $\xi _{\mathbf{k}-\mathbf{q}}\approx \xi _{%
\mathbf{k}}-\mathbf{v}_{\mathbf{k}}\cdot \mathbf{q}$ and
simplifies the analysis without changing the results
qualitatively. As long as $J\ll E_{F},$ we have $\left|
\mathbf{v}_{\mathbf{k}}\right| \approx k_{F}/m_{ab}\approx v_{F},$
where $k_{F}$ is the radius of the cylinder in Fig. 1 for $J=0$.
Despite the fact that the electron velocity does have a small
component along the $c$-axis, its in-plane component is large (cf.
Fig. 1). Since it is the magnitude of $\mathbf{v}_{\mathbf{k}}$
that controls the Migdal's approximation, the problem reduces to
the interaction of \emph{fast }2D electrons with \emph{slow }3D
phonons. With these simplifications, we find
\begin{subequations}
\begin{eqnarray}
\text{Re}\Sigma ^{R}(\epsilon ,\hat{n}) &=&-\frac{1}{4}\frac{\zeta }{1-\zeta
}\left( \frac{k_{D}}{k_{F}}\right) ^{2}\epsilon ;  \label{resigma} \\
\text{ Im}\Sigma ^{R}(\epsilon ,\hat{n}) &=&-\frac{\zeta }{12(1-\zeta )^{2}}%
\frac{\left| \epsilon \right| ^{3}}{\omega _{D}^{2}}\,,  \label{imsigma}
\end{eqnarray}
\end{subequations}
where $\zeta =\nu g^{2}$ is a dimensionless coupling constant and $\omega
_{D}=sk_{D}.$ We see that, despite the strong anisotropy, the self-energy
remains local, i.e., independent of $\xi _{\mathbf{k}}$.

Vertex renormalization leads to two types of corrections to the
self-energy: those that are proportional to the Migdal's parameter
($s/v_{F}$) and those that are proportional to 
$ms^{2}/\epsilon$. The second type of corrections invalidates the Migdal's
theory for temperatures below $ms^{2}$, which is about 1 K in a
typical metal. For metals with anisotropic spectrum the existence
of such a scale is potentially dangerous, since it is not obvious
which of the masses (light or heavy) defines this scale. We find
that the in-plane mass
($m_{ab}$) controls the vertex renormalization
for the nearly cylindrical Fermi surface. This shows that
the Migdal theory for layered metals has the same range of
applicability as for isotropic metals \cite{divergence}.

The rest of the derivation proceeds in the same way as for the
isotropic case \cite{rammer}, and the resulting Boltzmann equation
assumes its standard form. Since no assumption about the relation
between $\tau _{\text{e-ph}}$ and the dwell time ($1/J$) has been
made, the conductivities obtained from the Boltzmann equation have
the same form regardless of whether $J\tau _{\text{e-ph}}$ is
large or small. In other words, there is no coherent-incoherent
crossover due to inelastic scattering in an anisotropic metal
\cite{polarons}.

The situation changes qualitatively if resonant impurities are present in
between the layers. Electrons that tunnel through such impurities are moving
with the speed controlled by the broadening of a resonant level, i.e., much
slower than speed of sound. For that reason they can not be treated within
the formalism outlined above and require a separate study.

To evaluate the resonant-impurity contribution to the conductivity, we assume that the
impurities are randomly distributed in space with density
$n_{\mathrm{imp}}$ whereas their energy levels uniformly
distributed over an interval $E_{b}$. The tunneling conductance of
a bilayer junction is
\begin{equation}
G=-e^{2}\int d\epsilon d\epsilon ^{\prime }W_{\epsilon ,\epsilon ^{\prime }}%
\bigg[\frac{\partial n_{\epsilon }}{\partial \epsilon }(1-n_{\epsilon
}^{\prime })+\frac{\partial n_{\epsilon }^{\prime }}{\partial \epsilon
^{\prime }}n_{\epsilon }\bigg],
\end{equation}
where $W_{\epsilon ,\epsilon ^{\prime }}$ is a transition probability per
unit time and $n_{\epsilon }$ is the Fermi function. To calculate $%
W_{\epsilon ,\epsilon ^{\prime }},$ we use the results of
Ref.\cite {Glazman_1988,wingreen} for the probability
of phonon-assisted tunneling   through a single impurity
\begin{widetext}
\begin{eqnarray}&&
\label{c1}
W_{\epsilon,\epsilon'}=\Gamma_{\rm L}\Gamma_{\rm R}\int_{-\infty}^{\infty}dt_1e^{it_1(\epsilon'-\epsilon)}
\int_0^{\infty} dt_2dt_3 e^{i(t_2-t_3)(\epsilon-\bar{\epsilon}_0)-\Gamma(t_2+t_3)} \\&&
\times\exp
\left(-\sum_q\frac{|\alpha_q|^2}{2\omega_q^2}
\bigg[
|1-e^{-it_3}+e^{it_1}\left(e^{-it_2}-1\right)|^2
\coth\left(\frac{\omega_q}{2T}\right)+
\bigg(
e^{-it_3}
+e^{it_2}+
e^{it_1}(e^{-it_2}-1)(1-e^{it_3})-c.c.
\bigg)
\bigg]\right),\nonumber
\end{eqnarray}
\end{widetext}
where $\alpha _{q}=-i\Lambda q/\sqrt{\rho \omega _{q}}$, $\Lambda$ is the
deformation-potential constant, $\Gamma
_{\mathrm{L}}$ and $\Gamma _{\mathrm{R}}$ are tunneling widths of
the resonant level, $\Gamma =\Gamma _{\mathrm{L}}+\Gamma _{\mathrm{R}},$ and $\bar{%
\epsilon _{0}}$ is the energy of a resonant level renormalized by
the electron-phonon interaction. In the limit of no electron-phonon interaction,
Eq.({\ref{c1}}) reproduces the well-known Breit-Wigner formula. From now on,
we consider a wide band of resonant levels: $E_{b}\gg T\gg \Gamma $.
Averaging Eq.(\ref{c1}) over spatial and energy positions of resonant
levels, one obtains
\begin{eqnarray}
&&
\!\!\!\!\!\!\!\sigma_{\mathrm{res}}\!=\!\sigma_{\mathrm{el}}\!
\int_{-\infty}^{\infty
}\!\!\!\!\!d\epsilon\! \bigg[\!1\!-\!\coth\left(\!\frac{\epsilon}{2T}\!\right)\!+\!\frac{\epsilon}{2T}
\!\frac{1}{\sinh^2\left(\!\frac{\epsilon}{2T}\!\right)\!}\!\bigg]\!\int_{-\infty}^{\infty}\!\!\!\!dt e^{it\epsilon\!-\!\lambda f(t)} \nonumber
\label{res_tun} \\&&
\!\!\!\!\!f(t)\!=\!\int_0^{\omega_D}\!\!\!d\omega\frac{\omega}{\omega_D^2}\bigg[\!
(1\!-\!\cos(\omega t))\coth\left(\!\frac{\omega}{2T}\!\right)\!+\!i\sin(\omega t)\bigg]. 
\label{sigmares}
\end{eqnarray}
Here $\sigma _{\mathrm{el}}$ is the conductivity due to elastic
resonant tunneling and $\lambda \equiv \Lambda ^{2}\omega
_{D}^{2}/\rho s^{5}\pi ^{2}$ is the dimensionless coupling
constant for localized  electrons.
In the absence of
electron-phonon interaction, $\sigma _{\mathrm{res}}$ is temperature
independent and given by $\nolinebreak{\sigma _{\mathrm{el}%
}\simeq \pi e^{2}\Gamma _{1}n_{\mathrm{imp}}a_{0}d/E_{b}}$\cite
{Larkin_Matveev}, where $a_{0}$ is the localization radius of a
resonant state and $\Gamma _{1}\simeq \epsilon _{0}e^{-d/a_{0}}$
is its typical width.
We note that the electron-phonon interaction is much stronger for localized
electrons than for band ones: $\lambda /\zeta \sim \left(
k_{F}d\right) (v_{F}/s)\gg 1.$ Since typically $\zeta \sim 1,$ one
needs to consider a non-perturbative regime of phonon-assisted
tunneling.  In that case, resonant tunneling  
is exponentially suppressed at $T=0$: $\sigma_{\mathrm res}(T=0)=\sigma_{\mathrm el}e^{-\lambda/2}$.
At finite $T$, we find
\begin{align}
 \sigma_{\mathrm res}=\sigma_{\mathrm el}\left\{
\begin{array}{l}
e^{-\lambda/2}\left(1+\frac{\pi^2\lambda}{3}\left(\frac{T}{\omega_D}\right)^2\right)\, , \,\,\,   T\ll\frac{\omega_D}{\sqrt{\lambda}} \,\, ,
\\
1-\frac{\lambda}{9}\frac{\omega_D}{T}\,\, ,\,\,\,  T\gg \lambda\omega_D.
\end{array}
\right.
\nonumber
\end{align}
As $T$
increases, $\sigma _{\text{res }}$ growth, resembling the zero-bias anomaly in disordered metals
and M\"{o}ssbauer effect.
At high temperatures ($T\gg \lambda\omega_{D}$)  $\sigma _{\text{res}}$
 approaches the non-interacting value ($\sigma _{el}$). The asymptotic regimes
 in the interval $\omega_D/\sqrt{\lambda}\ll T\ll \lambda\omega_D$ can also be studied but we will
 not pause for this here.
 Notice that, in contrast
to the phenomenological model of Ref.\cite{levin}, there is no
simple relation between the $T$-dependences of $\sigma^B_c$ and
$\sigma_{{\rm res}}$.

To compare our model with the experiment, we extract $\sigma
_{c}^{B}$ from the low-temperature (between 10 and 50 K) $c$-axis
resistivity of Sr$_{2}$RuO$_{4}$ and extrapolate it to higher temperatures \cite{srruo}.
The resonant part of the conductivity is calculated numerically using Eq.(\ref{sigmares}).  The fit to the data for ${%
\sigma _{el}=43\cdot 10^{3}\,\Omega ^{-1}}$ cm$^{-1}$, $\omega _{D}=41$ K and $%
\lambda =16$ is shown in Fig. 2. The agreement between the theory
and experiment is quite good and the values of the fitting
parameters are reasonable. An immediate consequence of our model
is the sample-to-sample variation of the $c$-axis conductivity.
Among the layered materials, the largest amount of data is
collected for graphite \cite{graphite}. Even within the group of
samples with comparable in-plane mobilities, the temperature of
the maximum in $\rho _{c}$ varies from 40K to 300 K \cite
{graphite,hebard_unpub}.

\begin{figure}[tbp]
\label{fig2}
\includegraphics[angle=0,width
=0.45\textwidth]{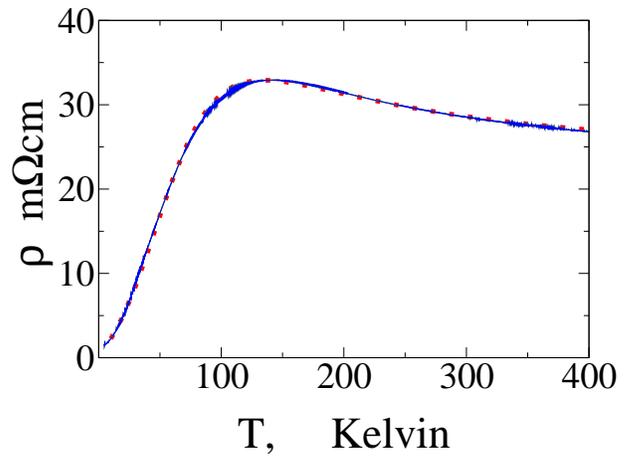}
\caption{$\rho_c$ vs
temperature. Solid: experimental data on Sr$_2$RuO$_4$; dashed:
fit into the phonon-assisted tunneling model in the non-perturbative
regime, Eq.(\ref{res_tun})}
\end{figure}

To conclude, we have shown that the Boltzmann equation and its
consequences are no less robust for anisotropic metals than they
are for isotropic ones. The only condition controlling the
validity of the Boltzmann equation is the large value of
$E_{F}\tau ,$ regardless of whether $\tau $ comes from elastic or
inelastic scattering. Out-of-plane localized states change the
$c$-axis transport radically while playing only minor role for the
in-plane one. While $\rho _{ab}$ remains metallic, an interplay
between phonon-assisted tunneling and conventional momentum
relaxation causes insulating or non-monotonic dependence of $\rho
_{c}$ on temperature. This model is in a good
agreement with the experimental data on Sr$_{2}$RuO$_{4}$.

This research was supported by NSF-DMR-0308377. We acknowledge
stimulating discussions with B. Altshuler, A. Chubukov, A. Hebard, S. Hill,
P. Hirschfeld, P. Littlewood, D. Khmelnistkii, N. Kumar, Yu. Makhlin, A.
Mirlin, M. Reizer, A. Schofield, S. Tongay, A.A. Varlamov, and P.
W\"{o}lfle. We are indebted to A. Hebard, A. Mackenzie, and S.
Tongay for making their data available to us.

\bigskip


\begin{thebibliography}{99}
\bibitem{divergence}  The self-energy in Eqs.(\ref{resigma},\ref{imsigma})
diverges at $\zeta =1.$ This divergence --also present for the isotropic
case -- results from the renormalization of the sound velocity and is an
artefact of the Fr{\"{o}}lich Hamiltonian. A divergence-free theory is
obtained by applying the adiabatic approximation to the coupled system of
electrons and ions \cite{schrieffer,Kagan,gelikman}.
\bibitem{polarons}  Migdal's theory also rules out models based entirely on
polaronic effects since polarons are stable only if a typical electron velocity
does not exceed the sound one.
\end{thebibliography}
\end{document}